\documentclass[twocolumn,showpacs,preprintnumbers,amsmath,amssymb]{revtex4}

\usepackage{graphicx}
\usepackage{dcolumn}
\usepackage{bbm,epsfig}

\newcommand{\be}{\begin{equation}}
\newcommand{\ee}{\end{equation}}
\newcommand{\ba}{\begin{array}}
\newcommand{\ea}{\end{array}}
\newcommand{\bqa}{\begin{eqnarray}}
\newcommand{\eqa}{\end{eqnarray}}

\newcommand{\bra}[1]{\ensuremath{\langle #1 |}}
\newcommand{\ket}[1]{\ensuremath{| #1 \rangle}}

\begin{document}

\title{Unravelling Entanglement}

\author{Andr\'e R.R. Carvalho, Marc Busse, Olivier Brodier, Carlos Viviescas,
  and Andreas Buchleitner}
\affiliation{Max-Planck-Institut f\"ur Physik komplexer Systeme,
N\"othnitzer Strasse 38, D-01187 Dresden}

\date{\today}

\begin{abstract}
We show that the time evolution of entanglement under incoherent environment
coupling can be faithfully recovered by monitoring the system according to a
suitable measurement scheme.  
\end{abstract}
\maketitle

Quantum entanglement, arguably the most non-classical feature of quantum
  mechanics, and considered as one of the key resources for quantum
  information processing, remains a puzzle for our intuition. General
  characterizations of the static entanglement and/or the separability
  properties of a given quantum state refer to the intricate geometry of
  tensor spaces, or to the operation of linear operations thereon. No
  general observable is known so far which would complement such essentially
  mathematical notions with a specific experimental measurement setup. Here we
  come up with a radically different, dynamical characterization of
  entanglement, through the  continuous 
  observation of a quantum system which evolves under incoherent coupling to
  an environment. We show that
  there is an optimal measurement strategy 
which provides an optimal decomposition of the time evolved, mixed
  system state, in the following sense: The entanglement of the time evolved
  mixed state is given by the average over the pure state entanglement of the
  single realisations of the stochastic time evolution. The weight of the
  different pure states is herein determined by their relative abundance
  during the detection process. Since arbitrary measurement
  prescriptions in general do {\em not} generate mixed state decompositions
  which are optimal in this sense, our finding implies that there are specific
  system observables which filter out the minimal amount of nonclassical
  correlations inscribed into a dynamically evolved, mixed quantum state.

Consider a bipartite 
quantum system composed of subsystems $A$ and $B$, interacting with 
its environment. Due to this coupling, an initial pure state $\ket{\Psi_0}$ of the
composite system will evolve
into a mixed state $\rho(t)$, in a way governed by 
the master equation $\dot \rho= \sum_k {\cal L}_k \rho$. The superoperators 
${\cal L}_k$ describe 
the effects of the environment on the system, and usually 
take the form~\cite{lind_76}
\be
\label{lindop}
{\cal L}_k  \rho= \frac{\Gamma_k}{2}\left(2\,J_k\,{\rho}\,J_k^\dagger -
J_k^\dagger\,J_k\,{\rho} -{\rho}\,J_k^\dagger\,J_k\right), 
\ee   
where the operators $J_k$ depend on the 
specific physical situation under study.

To extract the time evolution of entanglement under this incoherent dynamics,
one solution is to evaluate a given 
entanglement
measure $M(\rho)$ for the solution $\rho(t)$, 
at 
all times $t$. 
One starts 
from one of
the known
pure state measures 
$M(\Psi)$ \cite{Wo98,floPR05},
together with a pure state decomposition of $\rho$, 
\be
\label{decomp}
\rho=\sum_ip_i\ket{\Psi_i}\bra{\Psi_i},
\ee
where the $p_i$ are the positive, normalized weights 
of each pure state $\ket{\Psi_i}$.
The most
naive 
generalization for a mixed state would then be to consider the average 
\be
\label{ave_M}
\overline{M}=\sum_i p_i M(\Psi_i)\, ,  
\ee
which, however, 
is not uniquely defined, since the
decomposition~(\ref{decomp}) is not unique. Therefore, the proper definition
of $M(\rho)$ is the infimum of all possible averages
$\overline{M}$~\cite{uhl00}. 
Finding this infimum
has two main drawbacks: (i) it 
turns into a hard numerical
problem for higher dimensional or multipartite systems, and, (ii) even for
bipartite qubits, where analytical solutions for some 
measures $M(\rho)$
are known, there is no obvious general interpretation of the optimal decomposition, 
which gives the
infimum, in physical terms. 

\begin{figure}[t]
\epsfig{file=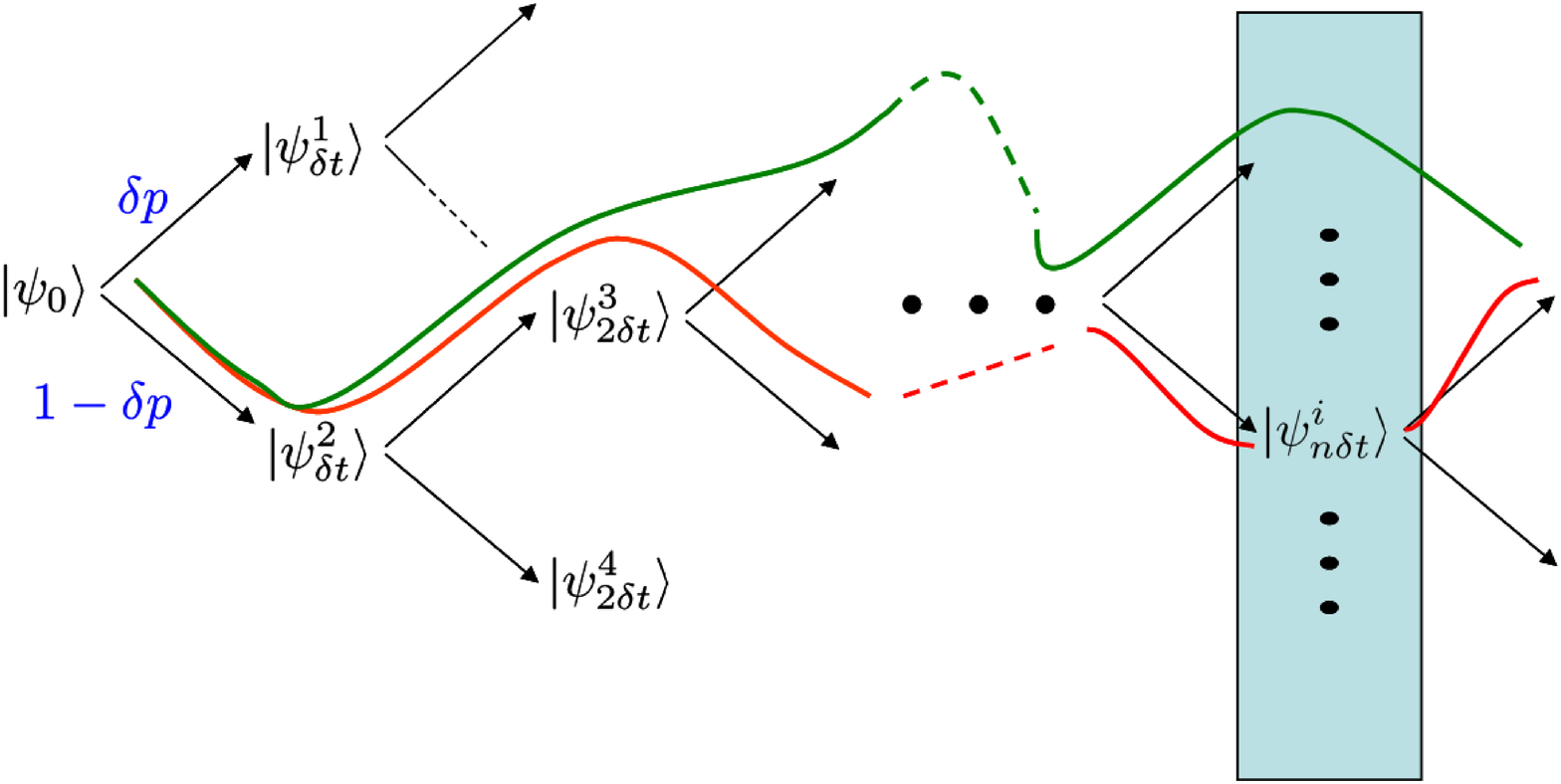,angle=0,width=0.45\textwidth}
\caption{Simplified schematic representation of stochastic time evolutions generated
  by continuous monitoring of the system dynamics through a jump operator
  $J$: During each time step of (short) duration $\delta t$, the system
  evolves with a certain probability $\delta p$ into the 
  state $\vert\Psi_{\delta t}^1\rangle$, and with probability ($1-\delta p$)
  into the state $\vert\Psi_{\delta t}^2\rangle$, depending on the result of
  the action of $J$ on the initial state $\vert\Psi_0\rangle$. Upon
  iterative application of this method, a large number of different states can
  be accessed, giving rise to different stochastic pure state trajectories. At
  a given time $t=n\delta t$, this process generates a pure state
  decomposition of the time evolved density matrix $\rho(t)$.
}
\label{fig1}
\end{figure}

Our approach here will be to avoid the direct use of the mixed state solutions
$\rho(t)$, by substituting them by physically motivated ensembles of pure
states. To do this, instead of solving the evolution equation for the density operator,
we will follow 
a stochastic time evolution
~\cite{carmichael,molmer93,molmer96} of 
the initially pure state. This
combines randomly occurring quantum jumps, defined by
the
action of the operators $J_k$ in Eq.~(\ref{lindop}), with periods of
continuous evolution, generated 
by a non-hermitian effective Hamiltonian.
The resulting 
quantum trajectories, see Fig.~\ref{fig1}, 
are known to provide,
after averaging over many independent realisations, the same
result as the master equation. 
Moreover, a single trajectory can be understood
as the experimentally accessible information through the 
continuous monitoring of the system~\cite{sauter86,wise93,molmer93,molmer96}.
In 
other words, one possible 
decomposition of the form (\ref{decomp}) 
is given by the set of 
the possible
outcomes $\ket{\Psi_i}$ of different runs of one and the same experiment, after a monitoring time
$t$.

Note, however, that the stochastic time evolution 
induced by the jump operators $J_k$ does 
not necessarily yield 
the optimal decomposition which minimises $\overline{M}$.
Yet, also the choice of the jump operators is not
unique: Indeed, the definition of new jumps as linear combinations of
the original ones, $\tilde J_k=\sum_i U_{ki}J_i$, with $U$ a unitary
matrix, or by the addition of a complex number $\mu$, 
$J^{\prime}_{k,\pm}=\left(\mu \pm J_k\right)/\sqrt{2}$, leaves
Eq.~(\ref{lindop}) unaltered. More generally, one can perform both changes
simultaneously, to
obtain the general jump 
operators $L_{k,\pm}=\left(\mu \pm \tilde J_k\right)/\sqrt{2}$. 

Since each such choice of jump operators produces 
a different {\it
  unravelling} of the same master equation (\ref{lindop}), i.e., 
a different decomposition of $\rho(t)$, we have to search for that 
type of monitoring which reproduces the desired optimal decomposition, upon
averaging over the different realisations at time $t$. 
We claim that such a measurement prescription can always be found, for a given
initial state, and given environment coupling.
While a rigorous mathematical proof of this conjecture
is still missing, we now present strong evidence in its favour,
for various decoherence
models.

We focus on the dynamics of bipartite two-levels systems, where the results
obtained by our method can readily be compared with known analytical
solutions~\cite{wot98,floPR05}. We assume that each subsystem interacts independently with
its own environment, and consider those situations where decoherence is induced
by dissipation (zero temperature reservoir), by dephasing, or by noise 
(infinite temperature reservoir). For simplicity, we set all decay rates
$\Gamma_k\equiv\Gamma$ in Eq.~(\ref{lindop}). We start with the general jump operators
$L_{k,\pm}$ as defined above, and
search for the ones which yield the optimal solution.  The first step in such
procedure is to follow all possible states $\vert\Psi_{\delta t}\rangle $ of the system, 
at a short time $\delta t$ after the initial time $t_0=0$, and to calculate
the average 
measure $\overline M(\delta t)$. 
Since the 
possible states $\vert \Psi_{\delta t}^k\rangle$ 
which can be reached after this short time interval are
given by the action of the $N$ jump operators, plus the state
$\vert\Psi_{\delta t}^{N+1}\rangle$ which results from continuous evolution
(see figure~\ref{fig1}),
the expression for the average measure is simply $\overline M({\delta
t})=(1-\sum_{k=1}^{N}\delta p_k)\,M(\Psi^{N+1}_{\delta t}) +
\sum_{k=1}^{N} \delta p_k
M(\Psi_{\delta t}^k)$, and can be minimised over the different possible
unravellings. This ensures that the
short time behaviour of the average measure is optimal.

In fact, for many initial states, such as the Bell states
$(\ket{00}+\ket{11})/\sqrt{2}$ and $(\ket{01}+\ket{10})/\sqrt{2}$ under
dephasing, or a state $\ket{\Psi_0}=\psi_{00} \ket{00}+\psi_{01}
\ket{01}+\psi_{10} \ket{10}$ coupled to a zero
temperature environment, the minimisation for the initial step can be
shown to be
not only a necessary but also a sufficient condition, for $\overline
M(t)$ to coincide with $M(\rho(t))$.
This is a striking feature,
since the optimisation needs to be performed only once, for the initial time
step, but leads to a faithful, optimal time evolution for arbitrary $t>t_0$. 

\begin{figure}[t]
\epsfig{file=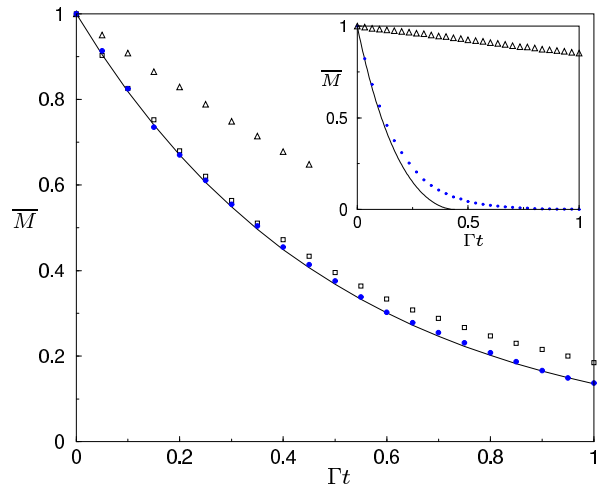,angle=0,width=0.45\textwidth}
\caption{Entanglement dynamics 
of a bipartite quantum state initially prepared in the maximally
  entangled state 
  $\ket{\Psi_0}=(\ket{00}+\ket{11})/\sqrt{2}$, under incoherent coupling to a zero temperature
  environment. $\overline M$ is here derived from the entanglement of
  formation \cite{Wo98}. The symbols result from an average over different realisations of the
  stochastic time evolution under the action of the general jump operators
  $L_{k,\pm}$, while the solid line represents the exact result derived from
  the integration of the master equation.   
Open squares ($\mu =1$) and filled
  circles ($\mu =3$) are obtained by
  progressively improved unravellings, with an optimal choice of the unitary
  $U$. Open triangles indicate the result of an arbitrarily chosen unravelling,
  with no optimization, neither of $U$ nor of $\mu$.}
\label{fig_decay}
\end{figure}

In other cases, however, the correct behaviour at short times is not
sufficient, and further search for the optimal unravelling is needed, though now in
a reduced parameter space (spanned by the parameters of
$L_{k,\pm}$). 
Such search is illustrated in Fig.~\ref{fig_decay},
for a Bell initial state $\ket{\Psi_0}=(\ket{00}+\ket{11})/\sqrt{2}$, coupled
to a zero temperature reservoir, and for $M(\Psi)$ given by the familiar
entanglement of formation \cite{Wo98}. 
The condition for the correct initial decay fixes
most of the parameters of the unitary matrix $U$, and enforces $\mu\ge
1/\sqrt{2}$. The open
squares and filled circles correspond to unravellings satisfying these
conditions for $\mu=1$, and $\mu=3$, respectively, while triangles are
obtained by a non-optimal choice of $U$ and $\mu$.
Although both values of $\mu$ induce the
correct behaviour 
for sufficiently short times, only the larger value of
$\mu$ gives good agreement for long times. The inset shows the situation for
the same initial state, though coupled to an infinite temperature
environment. The unravelling is chosen to give the correct initial slope, but the
agreement is also very good for the asymptotics. The only discrepancy persists
at intermediate times, where the exact solution for $M(\rho(t))$ exhibits a
discontinuous derivative. 

\begin{figure}
\epsfig{file=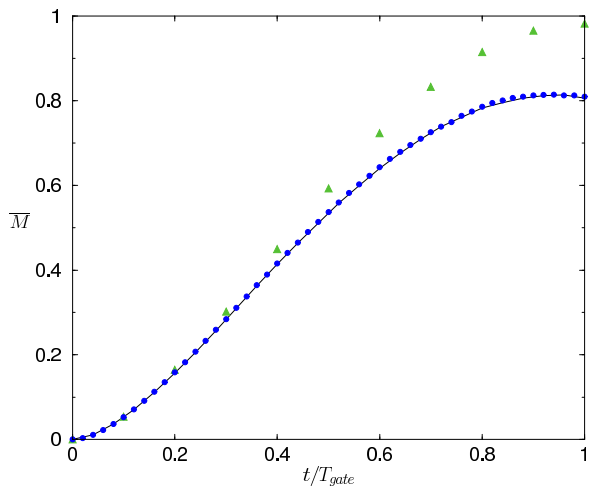,angle=0,width=0.45\textwidth}
\caption{Time evolution of the stochastically generated entanglement of
  formation (which we use here for the explicit evaluation of $\overline M$), 
during the application of a CNOT gate in a two qubit system coupled to a
  dephasing environment. Adequate choice of the jump operators $L_{k,\pm}$ leads
  to perfect agreement (circles) with the exact result  (solid line) during the gate
  time. A non-optimal choice of unravelling leads to large deviations from the exact solution.} 
\label{fig_CNOT}
\end{figure}

We can allow for more general dynamics, with a nontrivial unitary evolution in
the presence of incoherent environment coupling. 
This is relevant 
when it comes to describe, for example, gate operations performed in open quantum
systems. To illustrate such a situation, we simulated the action of a CNOT
gate, in a system coupled to a dephasing reservoir as shown in
Fig.~\ref{fig_CNOT}. 
Starting with a separable initial state
$\ket{\Psi_0}=(\vert 00\rangle+\vert 10\rangle )/\sqrt{2}$,
the gate would generate the maximally entangled state
$\ket{\Psi}=(\vert 00\rangle+\vert 11\rangle )/\sqrt{2}$, after a time $T_{\rm
  gate}$, chosen to be five times 
faster than the decoherence time scale $1/\Gamma$ (see Eq.~(\ref{lindop})). 
However, due to dephasing, the entanglement at $T_{\rm gate}$ will
not be maximal, since the target state preparation will not be perfect.
As in our above examples one can find an 
unravelling which generates an optimal pure state decomposition, in excellent 
agreement with the exact entanglement evolution obtained from the solution
$\rho(t)$. 
Once again, a non-optimal 
unravelling suggests a very different entanglement dynamics, with apparently
almost perfect gate performance -- in pronounced contrast to the correct
result. 

\begin{figure}[t]
\epsfig{file=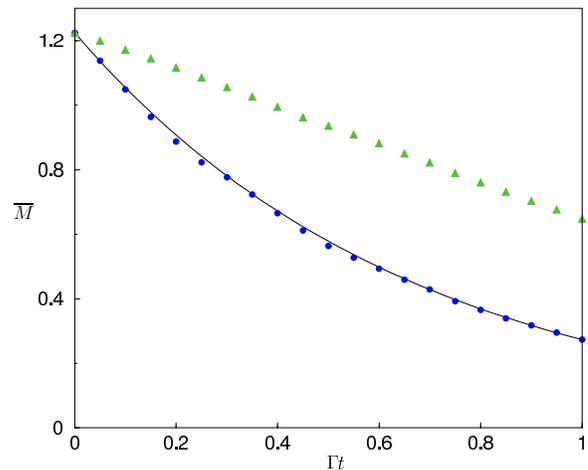,angle=0,width=0.45\textwidth}
\caption{Multipartite entanglement decay for a $\rm GHZ$ state under dephasing
  dynamics. Filled circles represent the tripartite concurrence
  $C_3(t)$ (substituted in the explicit evaluation of $\overline
  M$)~\cite{arrc_mpd} 
obtained after averaging over $1000$ 
  realisations of the unravelling defined by the jump operators 
  $J_k=\sigma^{(k)}_+ \,\sigma^{(k)}_{-}$ ($k=1,2,3$) (which actually coincide
  with the coupling operators in the dephasing master
  equation), in very good agreement with the exact solution indicated by the solid
  line. A non-optimal unravelling (open circles) considerably overestimates
  the actual mixed state entanglement of the time evolved state.
} 
\label{fig_mpd}
\end{figure}

Finally, let us consider the example of a tripartite qubit system, initially
prepared in a $\rm GHZ$ or in a $\rm W$ state, and in contact with dephasing
or zero temperature environments, respectively. In both cases, the optimal
unravelling produces excellent agreement with the exact result for the
tripartite concurrence $C_3(t)$ \cite{arrc_mpd,floPR05}, as illustrated
in Fig.~\ref{fig_mpd} for a $\rm GHZ$ initial state subject to dephasing. 
In
this exemplary case, the optimal detection is provided by the original jump
operators $J_k=\sigma^{(k)}_+ \,\sigma^{(k)}_{-}$, with $k\in
\{1,2,3\}$. Again, a non-optimal unravelling leads to very different predictions.

To conclude, we have shown that mixed state entanglement can be unraveled by a
suitable measurement prescription. The latter can be derived from the initial state
of the composite quantum system under study, together with the specific type of
environment coupling. There is thus no ambiguity (which was suggested in
\cite{nha04}) in the definition of mixed state entanglement with
respect to different unravellings: the minimal nonclassical correlations
needed to characterize a dynamically evolving quantum state are idenitified by
an optimal (in general not unique) monitoring prescription, and any in this sense
non-optimal unravelling provides a non-optimal estimate thereof.

Surprisingly, in all examples considered, we so far always succeeded to come up with a {\em
  time-independent} optimal unravelling, i.e., the monitoring strategy does
not need to be readapted during the incoherent time evolution. This suggests a
kind of uniform convergence of the stochastic average towards the exact
solution (which follows from the evaluation of the convex roof of a given
entanglement measure on the time evolved density matrix), for the optimal
unravelling, and implies a subtle relation between the temporal evolution of
  entanglement and the ensemble of pure states which can be reached during
  the single realisations generated by the optimal jump operators. 

We thank Florian Mintert, Daniel Est\`eve, Armin Uhlmann, and Robin Hudson for
entertaining discussions and useful suggestions.

\bibliography{quantinfo}

\end{document}